%% file: arxiv_August2026_1_.tex
\documentclass[11pt]{article}

\usepackage[letterpaper,margin=1in]{geometry}

\usepackage[T1]{fontenc}
\usepackage[utf8]{inputenc}
\usepackage{lmodern}
\usepackage{microtype}
\usepackage[hyphens]{url}
\usepackage[hidelinks]{hyperref}
\usepackage{graphicx}
\usepackage{caption}
\usepackage{booktabs}
\usepackage{array}

\usepackage{natbib}
\let\cite\citep
\usepackage{algorithm}
\usepackage{algorithmic}

\usepackage{newfloat}
\usepackage{listings}

\DeclareCaptionStyle{ruled}{
  labelfont=normalfont,
  labelsep=colon,
  strut=off
}

\floatstyle{ruled}
\newfloat{listing}{tb}{lst}{}
\floatname{listing}{Listing}

\usepackage{amsmath}
\usepackage{amsthm}
\usepackage{amssymb}
\usepackage{thm-restate}

\newtheorem{theorem}{Theorem}

\newtheorem{corollary}{Corollary}

\newtheorem{lemma}{Lemma}

\newtheorem{claim}{Claim}

\newcommand{\R}{\mathbb{R}}

\DeclareMathOperator*{\wdisc}{\mathrm{wdisc}}
\DeclareMathOperator*{\discpe}{\mathrm{disc}_{\mathrm{PE}}}

\title{
  Bad News for Couples:\\
  Bounds for Fair Division of Indivisible Items among Groups
}

\author{
  Max Dupré la Tour\\
  \small RIKEN Center for Advanced Intelligence Project\\
  \small University of Tokyo\\
  \small \texttt{maxduprelatour@gmail.com}
}

\date{}

\begin{document}

\maketitle

\begin{abstract}
We consider the problem of fairly allocating indivisible items to couples, where each couple consists of two agents with distinct additive valuations. We show that there exist binary instances with $n$ agents partitioned into $n/2$ couples for which envy-freeness up to $\Omega(\sqrt{n})$ items cannot be guaranteed. More generally, in the group-allocation model with $n$ agents partitioned into $k$ groups, we construct binary instances for which envy-freeness up to $\Omega(\sqrt{n-k})$ items cannot be guaranteed.
This matches the $O(\sqrt{n})$ upper bound of \citet{Manurangsi_ImprovedBounds} in all regimes except when $n-k \ll n$, that is, when most agents form singleton groups and only a sublinear number of agents belong to groups of size at least two. This result is somewhat surprising, as that upper bound was conjectured not to be tight for instances consisting only of small groups, such as couples.

We complement our lower bound with improved upper bounds for the remaining sparse regime. For prime-power $k$, we prove an upper bound of $O(\min\{\sqrt{(n-k)\log k},n-k\})$. For arbitrary $k$, the bound incurs an additional factor of $O(\frac{\log m}{\log k})$ where $m$ is the number of items. The result follows from a more general theorem that simultaneously guarantees approximate envy-freeness with respect to agents’ subjective valuations and approximate equality with respect to multiple consensus valuations.
\end{abstract}

\section{Introduction}
Resource allocation problems arise whenever limited resources must be distributed among multiple agents. Examples include allocating course seats, laboratory equipment, or administrative tasks. Despite their diversity, these settings share a common challenge: allocating resources in a way that is perceived as fair. The study of fair division addresses this challenge and has received attention from several disciplines, including economics, theoretical computer science, computational social choice, and multi-agent systems~\cite{brams1996fair,moulin2004fair,bouveret2016fair,aziz2020developments,walsh2020fair}.

We focus on the allocation of indivisible resources among agents with heterogeneous preferences~\cite{AmanatidisABFLMVW2023}. In the standard model, each agent has a valuation function over bundles of items. The goal is to partition the items into bundles, assign one bundle to each agent, and satisfy a desired fairness criterion. A central criterion is envy-freeness, which requires every agent to weakly prefer its own bundle to that of any other agent. For indivisible resources, exact envy-freeness may fail to exist, motivating approximate relaxations. A prominent relaxation is envy-freeness up to $c$ items (EF$c$): for any pair of agents, any envy can be eliminated by removing at most $c$ items from the envied bundle. In the standard model with monotone valuations, EF1 allocations are guaranteed to exist~\cite{LiptonEf1}, providing a baseline for many subsequent results.

This baseline can fail in extensions of the standard model. We focus on one such extension: allocation to groups rather than individuals. While most fair-division results concern individual agents, many applications involve small groups, such as couples or families, that receive a shared bundle even though their members may have different preferences. In this group-allocation model, bundles are assigned to groups, but fairness is evaluated at the level of individual agents, making guarantees more delicate.

\citet{Manurangsi_ImprovedBounds} initiated the systematic study of this model with $k$ groups and $n$ agents, where group sizes satisfy $n_1 \ge n_2 \ge \dots \ge n_k$ and $n=\sum_i n_i$. Using multicolor discrepancy results of \citet{MulticolorDoerr}, they showed that an allocation always exists that is envy-free up to $O(\sqrt{n})$ items and gave instances for which no allocation is envy-free up to $\Omega(\sqrt{n_1/k^3})$ items. For a constant number of large groups, these bounds are tight up to constant factors.

The dependence on $k$ in the lower bound was subsequently improved, first by \citet{caragiannis2025newlowerboundmulticolor} and later by \citet{ManurangsiM2026}, who strengthened the lower bound to $\Omega(\sqrt{n_1})$. A key ingredient in the latter result is a tight $\Omega(\sqrt{n})$ multicolor discrepancy lower bound that is independent of the number of colors.

For bounded group sizes, however, an intriguing gap remained: while the general upper bound is $O(\sqrt{n})$, the known lower bounds were only $\Omega(1)$. At the other extreme, when $n_1=\dots=n_k=1$, the model reduces to the
standard individual-agent setting and an EF$1$ allocation is always guaranteed~\cite{LiptonEf1}. This contrast led \citet{ManurangsiM2026} to suggest that sharper guarantees might be possible when groups are small.

\paragraph{Lower bound.} We show that this hope is unfounded: the worst-case behavior already arises for couples, where every group has size two. More generally, we prove a lower bound governed by the quantity $n-k$, the number of ``extra'' agents beyond one per group.

\begin{restatable}{theorem}{mainlb}\label{thm:main}
Let $n$ agents be partitioned into $k\geq 2$ non-empty groups of sizes $n_1,\dots,n_k$ with $\sum_{i=1}^k n_i=n$, and assume $n \geq k+4$.
There exist instances with binary additive valuations such that if an allocation is EF$c$, then
\[
c \ge  \frac{1}{96}
   \sqrt{\frac{n-k}{2}-1} =\Omega\!\left(\sqrt{n-k}\right).
\]
\end{restatable}

In particular, when every group has size two, we have $n-k=n/2$, yielding an $\Omega(\sqrt{n})$ lower bound. This result resolves open questions raised in several recent papers:~\cite{Bu, ManurangsiM2026, gölz2025fairdivisioncouplessmall}. Our proof proceeds via a simple reduction from the $p$-weighted discrepancy lower bound of \citet{ManurangsiM2026}.

\paragraph{Proof sketch of the lower bound.}
We sketch the proof for the special case of couples. Within each couple, the two agents have binary valuations and complementary preferences. Every item is valued by exactly one of the two partners. This construction is natural because it makes the two partners pull in opposite directions. As a result, it is difficult for one shared bundle to satisfy both agents.

More concretely, suppose one partner values exactly the items in $S$, and the other values exactly the items in $\overline S$. For any bundle, the two partners' values sum to the size of that bundle. Thus, the EF$c$ constraints for the two partners can be added to show that no other bundle can be much larger than their own bundle. Applying the same argument to every couple shows that all bundle sizes must be within an additive $O(c)$ of one another.

Now consider the number of $S$-items in each bundle. If another group receives too many items from $S$, then the first partner tends to envy that group. If it receives too few items from $S$, then, because bundle sizes are nearly balanced, it must receive many items from $\overline S$, so the second partner tends to envy it. Consequently, satisfying EF$c$ for both partners forces every group bundle to contain approximately $|S|/k$ items from $S$, up to an additive $O(c)$ error. This directly links approximate envy-freeness to $p$-weighted discrepancy.

\paragraph{Upper bound.}
The lower bound above shows that the general \(O(\sqrt n)\) guarantee cannot
be improved when \(n-k=\Theta(n)\). This leaves the sparse-group regime $n-k\ll n$ as the only regime in which one
can hope for a better general guarantee. In this setting, most groups are singletons and only a sublinear number of agents belong to groups of size at least $2$. Our second result gives a better guarantee in this regime.

Our upper bound is proved through a stronger auxiliary model. Each group has
one private agent, whose fairness constraints are imposed only for that group.
In addition, there are $t$ consensus valuations that are copied into every
group; these valuations must therefore be approximately balanced across all
group bundles.

\begin{restatable}{theorem}{mainub}\label{thm:mainub}
Let \(k \ge 2\) and \(t \ge 1\). Let 
\(\mathcal{V}= \{v_1,\dots,v_k\}\) and \(\mathcal{U} = \{u_1,\dots,u_t\}\) be two sets of non-negative additive valuations and consider the group-allocation
instance with \(k\) groups, where group \(i\) contains the private agent
\(v_i\) and the consensus agents \(u_1,\dots,u_t\). Then:
\begin{itemize}
    \item If \(k\) is a prime power, there exists an EF\(c\) allocation with
    \[
    c =
    O\!\left(
    \min\left\{
    \sqrt{t \log k},\,
    t
    \right\}\right).
    \]

    \item For arbitrary \(k\), there exists an EF\(c\) allocation with the same
    bound up to an additional factor of \(\frac{\log m}{\log k}\) where $m \geq k$ is the number of items; that is,
    \[
    c =
    O\!\left(
    \min\left\{
    \sqrt{t \log k},\,
    t
    \right\}
    \cdot \frac{\log m}{\log k}
    \right).
    \]
\end{itemize}
\end{restatable}

The proof is nonconstructive. It uses a
topological existence argument, and we do not know whether such an allocation
can be found in polynomial time.

\paragraph{Proof sketch of the upper bound.}

When there are only consensus agents and no private agents, the problem is
exactly the consensus \(1/k\)-division up to \(c\) items problem of
\citet{Manurangsi_ImprovedBounds}. They prove an \(O(\sqrt{t})\) upper
bound with a reduction to multicolor discrepancy. 

Our approach follows a similar route. We reduce the problem to a variant of
multicolor discrepancy in which, in addition to the usual balance constraints,
each color is subject to an approximate envy-freeness requirement induced by
the corresponding private valuation. We then establish discrepancy bounds for
this variant to prove Theorem~\ref{thm:mainub}.

We begin with the natural continuous relaxation, in which items may be
allocated fractionally. We require the \(k\) fractional bundles to be exactly
balanced with respect to every consensus valuation
\(u_1,\dots,u_t\), and we require the resulting fractional allocation to be
exactly envy-free with respect to the private valuations
\(v_1,\dots,v_k\). By placing
the items on a line and applying the equicardinal necklace-splitting theorem
of \citet{jojic2021splitting}, we obtain a fractional allocation in this
polytope for which each bundle contains only \(O(t)\) fractionally allocated
items.

The necklace-splitting theorem used here applies when \(k\) is a prime power.
In this case, assigning the fractional items integrally in an arbitrary manner
incurs an \(O(t)\) error. Alternatively, randomized rounding, together with
the concentration inequalities and a union bound, gives an
\(O(\sqrt{t\log k})\) error. Thus, for prime-power \(k\), we obtain the bound
\[
    O\!\left(\min\left\{t,\sqrt{t\log k}\right\}\right).
\]

When \(k\) is not a prime power, let \(K\geq k\) be the smallest
prime power to which the preceding argument applies, and introduce
\(K-k\) dummy bundles. We apply the \(K\)-bundle construction, retain the
items assigned to the original \(k\) bundles, and recursively reallocate the
items assigned to the dummy bundles. Each recursive level incurs an additive
error of $O\!\left(\min\left\{t,\sqrt{t\log K}\right\}\right)$.

The prime-gap bound~\cite{baker2001difference} guarantees that we may choose $K = k + O(k^{0.525})$. Consequently, $\log K = \Theta(\log k)$ and $\frac{K - k}{K} = O(k^{-0.475})$. After one recursive level only an
\(O(k^{-0.475})\) fraction of the items remains to be allocated. It follows that $r=O\!\left(\frac{\log m}{\log k}\right)$
recursive levels suffice.




\medskip 
We now return to the original group-allocation model. Given \(k\) groups and
\(n\) agents, set \(t=n-k\). Choose one representative from each group as the
private agent, and treat all remaining agents as consensus agents. That is, we
copy each remaining agent into every group. This only strengthens the fairness
requirements. Hence any EF\(c\) allocation for the copied instance is also
EF\(c\) for the original group instance. We obtain the
following guarantee.

\begin{corollary}\label{cor:groupub}
Let \(n>k\ge 2\), and let \(n\) agents be partitioned into \(k\) groups. For
arbitrary additive valuations, if \(k\) is a prime power, there exists an
EF\(c\) allocation with
\[
c =
O\!\left(
\min\left\{
\sqrt{(n-k)\log k},\;
n-k
\right\}
\right).
\]
If \(k\) is arbitrary, the same bound holds up to an additional factor of
\(\frac{\log m}{\log k}\), where $m\geq k$ is the number of items.
\end{corollary}

Combining Corollary~\ref{cor:groupub} with the general $O(\sqrt n)$ upper bound of Manurangsi and Suksompong, these results show that $n-k$, the number of agents beyond one representative per group, is the key parameter governing the difficulty of achieving EF$c$ allocations in the group-allocation model. When $k$ is a prime power, we obtain the following picture.

\begin{itemize}
    \item If \(n-k=\Theta(n)\), the upper and lower bounds match up to constant
    factors: the optimal guarantee is
    \(\Theta(\sqrt n)=\Theta(\sqrt{n-k})\).

    \item If \(n-k=o(n)\), there is only a logarithmic gap between the
    \(O(\sqrt{(n-k)\log k})\) upper bound and the
    \(\Omega(\sqrt{n-k})\) lower bound.

    \item In the very sparse regime \(n-k<\log k\), the stronger upper bound
    \(O(n-k)\) improves on \(O(\sqrt{(n-k)\log k})\). In particular, this
    yields an EF\(c\) allocation with constant \(c\) when one group has
    constant size and all other groups are singletons.
\end{itemize}

The prime-power assumption on \(k\) is somewhat unsatisfactory: without it, our bound
loses a factor \(\log m/\log k\). Similar assumptions appear in topological
fair-division results~\cite{avvakumov2021envy,igarashi2025envy}, and is known to be
necessary in some cake-cutting settings~\cite{avvakumov2021envy}.

Removing this assumption, and more generally understanding the corresponding variant of multicolor discrepancy with additional private valuations, is an interesting direction for future work. Beyond proving existential guarantees, it would be particularly interesting to develop efficient algorithms achieving them.

\subsection{Related work}

There is a rich literature on fair division for groups, spanning a range of settings and fairness notions. Examples include cake cutting~\cite{Segal-Halevi,SegalHalevi2020HowTC,Segal3}, maximin share guarantees~\cite{ManurangsiOrdinal,SUKSOMPONGMaximin}, asymptotic fairness under random valuations~\cite{MANURANGSI2017Asymptotic}, and models in which groups are not fixed but can be chosen to facilitate fairness~\cite{KYROPOULOUsuksompong,gölz2025fairallocationindivisiblegoods}. We next summarize results most closely related to our setting; each of the cited papers contains many additional contributions beyond those we mention here.

In the two-group setting, \citet{SEGALHALEVI2019103167} gave the first general existence guarantee, showing that for $n$ agents split into two groups, an allocation satisfying EF$n$ is guaranteed to exist. The first nontrivial lower bound, $\Omega(\log n)$, was proved by \citet{KYROPOULOUsuksompong}, who also showed that EF$1$ may not exist for two groups of three agents. Using multicolor discrepancy theory, these bounds were later sharpened in \cite{Manurangsi_ImprovedBounds,caragiannis2025newlowerboundmulticolor,ManurangsiM2026}: they currently stand at $O(\sqrt{n})$ for the upper bound and $\Omega(\sqrt{n_1})$ for the lower bound, where $n_1$ is the size of the largest group. Discrepancy-based techniques have also been partially extended beyond additive valuations~\cite{DUPRELATOUR,Hollender}, yielding an $O(\sqrt{n\log n})$ upper bound for EF$c$ with $n$ agents partitioned into two groups.

\citet{Bu} give an algorithm producing an EF$1$ allocation for a couples model in which one member of each couple is a copy of the same agent. They also prove existence of an EF$1$ allocation for two couples, and ask whether EF$1$ continues to hold for more couples, or more generally what is the smallest $c$ such that EF$c$ is always guaranteed.

\citet{gölz2025fairdivisioncouplessmall} provide another proof of EF$1$ existence for two couples, and further show that EF$1$ may fail for three couples.

\citet{kawase2025simultaneouslyfairallocationindivisible} study a variant in which the same set of items must be removed to eliminate the envy of all members of a group.

Our upper bound simultaneously guarantees approximate envy-freeness with respect to agents' subjective valuations and approximate equality with respect to multiple consensus valuations. This model generalizes one of the couple settings of \citet{Bu}, in which one member of each couple is a copy of the same agent. \citet{barman2025fair} study a closely related model in which allocations are evaluated according to both agents' subjective valuations and a common market valuation. Our framework extends this model by allowing multiple common valuations simultaneously. More broadly, our results contribute to the study of fair division under additional constraints; see the survey of \citet{suksompong2021constraints}.

\section{Preliminaries}
Let $M$ be a set of $m$ indivisible items. There are $n$ agents partitioned into $k$ groups
$G_1,\dots,G_k$ of sizes $n_1,\dots,n_k$ such that $\sum_{g=1}^k n_g=n$.
Each agent has an additive valuation $v:2^M\to\R_{\geq0}$. A valuation is binary additive if
each item has value in $\{0,1\}$.

An allocation is a partition of the items into $k$ bundles
$
M=A_1\sqcup\cdots\sqcup A_k,
$ 
where group $G_g$ receives bundle $A_g$.

Fix an integer $c\ge 0$. An allocation is EF$c$ if for every pair of groups $i,j\in[k]$
and every agent $a\in G_i$, there exists a set $R\subseteq A_j$ with $|R|\le c$ such that
\[
v_a(A_i)\ \ge\ v_a(A_j\setminus R).
\]
\section{Proof of the lower bound}

\mainlb*

We use the notion of $p$-weighted discrepancy. Given a set $\mathcal{V}=\{v_1,\dots,v_t\}$ of additive
valuation functions and a parameter $p\in(0,1)$, define
\[\textstyle
\wdisc_p(\mathcal{V})
:=\displaystyle\min_{A\subseteq M}\ \max_{i\in[t]}\Bigl|p\cdot v_i(M)-v_i(A)\Bigr|.
\]
We use the following lower bound.

\begin{theorem}[\citealp{ManurangsiM2026}]\label{thm:manurangsimekaLB}
For every $p\in(0,1)$ and $t\in $, there exist a set $M$ of items and a collection
$\mathcal{V}=\{v_1,\dots,v_t\}$ of $t \in \mathbb{Z}_{>0}$ binary additive valuation functions such that
\[\textstyle
\wdisc_p(\mathcal{V})\ \ge\ \frac{\sqrt{t-1}}{16}.
\]
\end{theorem}

The theorem follows from combining Theorem~\ref{thm:manurangsimekaLB} with the following lemma.

\begin{lemma}\label{lem:main-group}
Let $\mathcal{S}=\{S_1,\dots,S_t\}$ be a family of $t$ subsets of $M$. For each $i\in[t]$, consider a couple (two agents) with valuations
\[
v_i^1(A):=|S_i\cap A| \qquad\text{and}\qquad v_i^2(A):=|\overline{S_i}\cap A|,
\]
where $\overline{S_i}:=M\setminus S_i$.
Assume these $t$ couples are placed into the $k$ non-empty groups so that each couple lies entirely in one group; let $g(i)\in[k]$ denote the group containing couple $i$.
Additionally, for every group $G_g$ that contains a leftover (unpaired) agent, assign that agent the counting valuation $u(A):=|A|$.

Assume there exists an allocation $M=A_1\sqcup\cdots\sqcup A_k$ that is EF$c$ for some integer $c\ge 0$ with respect to all these valuations.
Let $\mathcal{V}=\{v_1^1,\dots,v_t^1\}$ denote the $t$ binary valuations $v_i^1(A):=|S_i\cap A|$.
Then $\wdisc_{1/k}(\mathcal{V})\ \le\ 6c$.

\end{lemma}
\begin{proof}
Let $M=A_1\sqcup\cdots\sqcup A_k$ be an EF$c$ allocation, and let $m:=|M|$.

The following claim is the standard consequence of EF$c$ for binary additive valuations.

\begin{claim}\label{cl:ef-ineq-group}
For every $i\in[t]$, every $h\in[k]$, and each $r\in\{1,2\}$,
\begin{equation}\label{eq:ef-ineq-group}
v_i^r(A_h)\ \le\ v_i^r(A_{g(i)})+c.
\end{equation}
Moreover, for every group $g$ that contains a counting agent $u$, and every $h\in[k]$,
\begin{equation}\label{eq:ef-ineq-counter}
|A_h|=u(A_h)\ \le\ u(A_g)+c=|A_g|+c.
\end{equation}
\end{claim}

The first step is to show that all bundle sizes are within $2c$ of each other.

\begin{claim}\label{cl:sizes-group}
For all $g,h\in[k]$,
\[
\bigl||A_g|-|A_h|\bigr|\ \le\ 2c.
\]
Consequently, for every $g\in[k]$,
\begin{equation}\label{eq:size-avg-group}
\frac{m}{k}-2c\ \le\ |A_g|\ \le\ \frac{m}{k}+2c.
\end{equation}
\end{claim}

\begin{proof}[Proof of Claim~\ref{cl:sizes-group}]
Note that every group contains either at least one of the couples or a leftover counting agent. Fix $g,h\in[k]$.
If group $G_g$ contains a counting agent $u$, then \eqref{eq:ef-ineq-counter} gives $|A_h|\le |A_g|+c$, which implies $|A_h|\le |A_g|+2c$.
Otherwise, group $G_g$ contains at least one couple; pick any couple $i$ with $g(i)=g$.
Apply \eqref{eq:ef-ineq-group} for $r=1,2$ and add:
\[
v_i^1(A_h)+v_i^2(A_h)\ \le\ v_i^1(A_g)+v_i^2(A_g)+2c.
\]
But for any set $A$,
\[
v_i^1(A)+v_i^2(A)=|S_i\cap A|+|\overline{S_i}\cap A|=|A|.
\]
Hence $|A_h|\le |A_g|+2c$.
Swapping the roles of $g$ and $h$ yields $|A_g|\le |A_h|+2c$, proving $\bigl||A_g|-|A_h|\bigr|\le 2c$.
Since $\sum_{g=1}^k |A_g|=m$, the average $m/k$ lies between $\min_g|A_g|$ and $\max_g|A_g|$, which implies \eqref{eq:size-avg-group}.
\end{proof}

We now show that every couple forces each bundle to contain approximately a $1/k$-fraction of its set.

\begin{claim}\label{cl:goal-group}
For every $i\in[t]$ and every $h\in[k]$,
\begin{equation}\label{eq:goal-group}
\left|\frac{|S_i|}{k}-|S_i\cap A_h|\right|\ \le\ 6c.
\end{equation}
\end{claim}

\begin{proof}[Proof of Claim~\ref{cl:goal-group}]
Fix $i\in[t]$ and $h\in[k]$.
Suppose, for contradiction, that
\begin{equation}\label{eq:contra-group}
\left|\frac{|S_i|}{k}-|S_i\cap A_h|\right|\ >\ 6c.
\end{equation}
There are two cases.

\noindent\emph{Case 1: $|S_i\cap A_h|<\frac{|S_i|}{k}-6c$.}
Then
\[
v_i^2(A_h)=|\overline{S_i}\cap A_h| =|A_h|-|S_i\cap A_h| >|A_h|-\frac{|S_i|}{k}+6c.
\]
Using the lower bound $|A_h|\ge \frac{m}{k}-2c$ from \eqref{eq:size-avg-group}, we obtain
\begin{align*}
v_i^2(A_h)
&> \left(\frac{m}{k}-2c\right)-\frac{|S_i|}{k}+6c \\
&= \frac{m-|S_i|}{k}+4c \\
&= \frac{|\overline{S_i}|}{k}+4c.
\end{align*}
By \eqref{eq:ef-ineq-group} applied to valuation $v_i^2$,
\[
v_i^2(A_{g(i)})\ \ge\ v_i^2(A_h)-c\ >\ \frac{|\overline{S_i}|}{k}+3c.
\]
Therefore,
\begin{align*}
v_i^1(A_{g(i)})
&= |S_i\cap A_{g(i)}| \\
&= |A_{g(i)}|-v_i^2(A_{g(i)}) \\
&< \left(\frac{m}{k}+2c\right)
   -\left(\frac{|\overline{S_i}|}{k}+3c\right) \\
&= \frac{|S_i|}{k}-c .
\end{align*}
Here we used the upper bound $|A_{g(i)}|\le \frac{m}{k}+2c$ from \eqref{eq:size-avg-group}.
On the other hand,
\[
\sum_{t'=1}^k v_i^1(A_{t'}) =\sum_{t'=1}^k |S_i\cap A_{t'}| =|S_i\cap M| =|S_i|.
\]
Hence the average value of $v_i^1(A_{t'})$ over $t'\in[k]$ is $|S_i|/k$, so there exists $\ell\in[k]$ with $v_i^1(A_\ell)\ge \frac{|S_i|}{k}.$
Combining with $v_i^1(A_{g(i)})<\frac{|S_i|}{k}-c$ gives $v_i^1(A_\ell)>v_i^1(A_{g(i)})+c$, contradicting \eqref{eq:ef-ineq-group} for $r=1$.
Thus Case~1 is impossible.

\noindent\emph{Case 2: $|S_i\cap A_h|>\frac{|S_i|}{k}+6c$.}
Then $v_i^1(A_h)=|S_i\cap A_h|>\frac{|S_i|}{k}+6c$.
By \eqref{eq:ef-ineq-group} applied to valuation $v_i^1$,
\[
v_i^1(A_{g(i)})\ \ge\ v_i^1(A_h)-c\ >\ \frac{|S_i|}{k}+5c.
\]
Therefore,
\begin{align*}
v_i^2(A_{g(i)})
&= |\overline{S_i}\cap A_{g(i)}| \\
&= |A_{g(i)}|-|S_i\cap A_{g(i)}| \\
&= |A_{g(i)}|-v_i^1(A_{g(i)}) \\
&< \left(\frac{m}{k}+2c\right)
   -\left(\frac{|S_i|}{k}+5c\right) \\
&= \frac{m-|S_i|}{k}-3c \\
&= \frac{|\overline{S_i}|}{k}-3c .
\end{align*}
Again we used~\eqref{eq:size-avg-group}.
As before,
\[
\sum_{t'=1}^k v_i^2(A_{t'})=\sum_{t'=1}^k |\overline{S_i}\cap A_{t'}|=|\overline{S_i}|,
\]
so there exists $\ell\in[k]$ with $v_i^2(A_\ell)\ge |\overline{S_i}|/k$.
Thus $v_i^2(A_\ell)>v_i^2(A_{g(i)})+3c$, contradicting \eqref{eq:ef-ineq-group} for $r=2$.
Therefore Case~2 is impossible.

Since both cases lead to contradictions, \eqref{eq:contra-group} is false, proving \eqref{eq:goal-group}.
\end{proof}

Finally, we upper bound $\wdisc_{1/k}(\mathcal{V})$ by choosing the subset $A_1$ in the definition:
\begin{align*}
\wdisc_{1/k}(\mathcal{V})
&=\min_{S\subseteq M}
  \max_{i\in[t]}
  \left|
  \frac{1}{k}v_i(M)-v_i(S)
  \right| \\
&\le
  \max_{i\in[t]}
  \left|
  \frac{|S_i|}{k}-|S_i\cap A_1|
  \right| \\
&\le 6c .
\end{align*}
where the last inequality is Claim~\ref{cl:goal-group}.
This completes the proof.
\end{proof}

\begin{proof}[Proof of Theorem~\ref{thm:main}]
Fix $k\ge 2$ and group sizes $n_1,\dots,n_k$ with $\sum_{g=1}^k n_g=n$. For each group \(g\), let \(t_g=\lfloor n_g/2\rfloor\) and
\(t=\sum_g t_g\). Then
\begin{equation}\label{eq:t-lb}
t=\sum_g\left\lfloor\frac{n_g}{2}\right\rfloor
\ge \sum_g\frac{n_g-1}{2}=\frac{n-k}{2}.
\end{equation}

Apply Theorem~\ref{thm:manurangsimekaLB} with $p=1/k$ and $t$ valuations.
We obtain a set of items $M$ and binary additive valuations $\mathcal{V}=\{v_1,\dots,v_t\}$ such that
\[
\textstyle \wdisc_{1/k}(\mathcal{V})\ \ge\ \displaystyle \frac{\sqrt{t-1}}{16}.
\]
Since the valuations are binary additive, for each $i\in[t]$ there exists $S_i\subseteq M$ such that $v_i(A)=|S_i\cap A|$.

We now construct the group instance.
In each group $G_g$, create $t_g$ couples, and assign them distinct indices from $[t]$.
Couple $i$ has two agents with valuations
\[
v_i^1(A):=|S_i\cap A| \qquad\text{and}\qquad v_i^2(A):=|\overline{S_i}\cap A|.
\]
If $n_g$ is odd, there is one remaining (unpaired) agent in $G_g$; assign that agent the counting valuation $u(A)=|A|$.

Suppose, toward contradiction, that there exists an EF$c$ allocation $M=A_1\sqcup\cdots\sqcup A_k$.
By Lemma~\ref{lem:main-group}, this implies
\[
\textstyle \wdisc_{1/k}(\mathcal{V})\ \le\ 6c.
\]
Combining with the lower bound gives
\[
6c\ \ge\ \frac{\sqrt{t-1}}{16} \qquad\Longrightarrow\qquad c\ \ge\ \frac{\sqrt{t-1}}{96}.
\]
Using \eqref{eq:t-lb}, we obtain
\begin{align*}
c
&\ge \frac{1}{96}
   \sqrt{\frac{n-k}{2}-1} \\
&= \Omega(\sqrt{n-k}) .
\end{align*}
\end{proof}

\section{Proof of the upper bound}

\mainub*
Following the approach of \citet{Manurangsi_ImprovedBounds}, we relate this question to the natural multicolor discrepancy variant.

Let $M$ be a set of $m$ indivisible items. Fix $k\ge 2$ and $t\ge 1$. Consider $\mathcal{V}=\{v_1,\ldots,v_k\}$ and $\mathcal{U}=\{u_1,\ldots,u_t\}$, where each valuation is additive and $v_i(g),u_\ell(g)\in[0,1]$ for all $g\in M$, $i\in[k]$, and $\ell\in[t]$. For an allocation $\mathcal{A}=(A_1,\dots,A_k)$, define
\begin{align*}
B(\mathcal{U},\mathcal{A})
&=
\max_{\ell\in[t],\,i,j\in[k]}
\left|u_\ell(A_i)-u_\ell(A_j)\right|, \\
P(\mathcal{V},\mathcal{A})
&=
\max_{i,j\in[k]}
\bigl(v_i(A_j)-v_i(A_i)\bigr)_+ .
\end{align*}
The term $B(\mathcal{U},\mathcal{A})$ is the usual multicolor balance error for the consensus valuations,
while $P(\mathcal{V},\mathcal{A})$ is the private-envy error. We define
\[
\discpe(\mathcal{V}, \mathcal{U},\mathcal{A})=\max\{B(\mathcal{U},\mathcal{A}),P(\mathcal{V},\mathcal{A})\}.
\]

The private-envy discrepancy of the instance is
\[
\discpe(\mathcal{V}, \mathcal{U})
=
\min_{\mathcal{A}}
\discpe(\mathcal{V}, \mathcal{U},\mathcal{A}),
\]
where the minimum is over all ordered partitions of $M$. 

Note that, as is often the case in the discrepancy literature, we define this variant of discrepancy only for families of valuations that assign each item a value in $[0,1]$, in order to avoid a normalizing factor in the bounds.

Our next lemma establishes a connection between private-envy discrepancy and EF$c$ guarantees in the group allocation setting of Theorem~\ref{thm:mainub}. Specifically, it shows that upper and lower bounds on $\discpe$ imply corresponding upper and lower bounds, respectively, on the EF$c$ guarantee for the group allocation instance of Theorem~\ref{thm:mainub}. Only the upper-bound implication is used in the proof of the theorem.

\begin{restatable}{lemma}{discrepancyreduction}\label{lem:discrepancy_reduction}
Consider an instance with a set $\mathcal{V}=\{v_1,\ldots,v_k\}$ of $k$ private valuations and a set $\mathcal{U}=\{u_1,\ldots,u_t\}$ of $t$ consensus valuations. Suppose that each valuation is additive and assigns each item a value in $[0,1]$. Let $c \geq 1$ be an integer such that $\discpe(\mathcal{V},\mathcal{U}) > c$. Then the group allocation instance with $\mathcal{V}$ as the set of private valuations and $\mathcal{U}$ as the set of consensus valuations; admits no EF$c$ allocation.

Conversely, let $f : \mathbb{Z}_{>0}^3 \to \mathbb{Z}_{>0}$ be such that, for every $(k,t,m)$, $f(k,t,m)$ upper bounds the private-envy discrepancy of every $[0,1]$-valued instance with $k$ private valuations, $t$ consensus valuations, and $m$ items. 

Then every group-allocation instance with arbitrary nonnegative additive valuations with $k$ private valuations and $t$ consensus valuations admits an EF$c$ allocation, where $c = 4f(k,2t,m)$.
\end{restatable}

The proof follows Section~3 of \citet{Manurangsi_ImprovedBounds}, the only new
ingredient is an initial round-robin phase using the private valuations.
The full proof is in the appendix. It remains to prove an upper bound on
\(\discpe\), which by Lemma~\ref{lem:discrepancy_reduction} implies
Theorem~\ref{thm:mainub}.
\medskip

A fractional \(k\)-allocation of a set \(M\) is a vector
\(x=(x^i_{g})_{i\in[k],g\in M}\) such that \(x^i_{g}\ge 0\) and
\(\sum_{i=1}^k x^i_{g}=1\) for every item \(g\).  We say that \(x\) is
\(q\)-sparse if, for every color \(i\in[k]\), at most \(q\) items satisfy
\(0<x^i_{g}<1\).

\begin{restatable}{lemma}{sparserounding}\label{lem:sparse_rounding}
There is a universal constant $C>0$ such that the following holds. Let $x$ be a $q$-sparse fractional $k$-allocation of $M$, and let $w_1,\ldots,w_d$ be additive valuations with $w_r(g)\in[0,1]$. Then there exists an integral allocation $\mathcal A=(A_1,\ldots,A_k)$ such that, for all $r\in[d]$ and $i\in[k]$,
$$
\left|w_r(A_i)-\sum_{g\in M}x_g^i w_r(g)\right|
\le
C\min\left\{q,\sqrt{q\log(kd)}\right\}.
$$
Moreover, the rounding is support-preserving: if $g\in A_i$, then $x_g^i>0$.
\end{restatable}

Any support-preserving rounding scheme guarantees an error of at most $q$. The second bound follows by applying a concentration inequality and then taking a union bound. See the appendix for the full proof.

To obtain a sparse fractional allocation to which the rounding lemma can be applied, we use the following necklace splitting result due to \citet{jojic2021splitting}. The original statement is more general, but we only
need the following weaker version in which each agent's preferences are induced
by a continuous probability measure on \([0,1]\).

\begin{theorem}[{\citealp[Theorem~6.14]{jojic2021splitting}}]\label{thm:jojic}
Let the number of agents \(k\geq 2\) be a prime power, and let \(t\geq 1\). Let
\(\mu_1,\ldots,\mu_t\) be continuous additive probability measures on
\([0,1]\), corresponding to the global balancedness conditions. For each
agent \(j\in[k]\), let \(\nu_j\) be a continuous additive probability measure
on \([0,1]\), defining agent \(j\)'s preferences.

Then \([0,1]\) can be cut at at most \((k-1)(t+1)\) points, and the resulting intervals allocated to bundles
\((A_1,\ldots,A_k)\) so that, after relabeling the bundles,
\[
\mu_\ell(A_i)=\frac{1}{k}
\qquad
(\ell\in[t],\ i\in[k]),
\]
and
\[
\nu_i(A_i)\ge \nu_i(A_j)
\qquad
(i,j\in[k]).
\]
Moreover, each \(A_i\) is a union of at most \(t+1\) intervals.
\end{theorem}

The next lemma uses this theorem to obtain a sparse fractional allocation.

\begin{lemma}\label{lem:sparse_fractional_allocation}
Let $k\ge 2$ be a prime power and let $t\ge 1$. Let $M$ be a set of items, and
let $\mathcal U=\{u_1,\ldots,u_t\}$ and $\mathcal V=\{v_1,\ldots,v_k\}$ be
additive valuations with item values in $[0,1]$. Then there exists a fractional
$k$-allocation $x$ such that, for all $\ell\in[t]$ and $i,j\in[k]$,
$$
\sum_{g\in M} x_g^i u_\ell(g)
=
\sum_{g\in M} x_g^j u_\ell(g),
$$
for all $i,j\in[k]$,
$$
\sum_{g\in M} x_g^i v_i(g)
\ge
\sum_{g\in M} x_g^j v_i(g),
$$
Moreover, $x$ can be chosen $O(t)$-sparse.
\end{lemma}
\begin{proof}
Order the items
arbitrarily and identify each $g\in M$ with a unit interval $I_g$. Each nonzero
valuation $w$ induces a non-atomic probability measure by assigning constant
density $w(g)$ on $I_g$ and normalizing by $w(M)$. In the edge case where $w$ is the $0$ valuation and $w(M) = 0$, choose any arbitrary non-atomic probability measure.

Apply Theorem~\ref{thm:jojic} to the consensus measures induced by
$u_1,\ldots,u_t$ and to the private measures induced by
$v_1,\ldots,v_k$. We obtain a partition $F_1,\ldots,F_k$ such that the
consensus measures are split equally and color $i$ is envy-free for $v_i$.

Set
$
x_g^i=|F_i\cap I_g|
\qquad (g\in M,\ i\in[k]).
$
Then $\sum_i x_g^i=1$, so $x$ is a fractional $k$-allocation. Equal splitting,
after undoing the normalizations, gives
$$
\sum_g x_g^i u_\ell(g)=\sum_g x_g^j u_\ell(g)
$$
for all $\ell\in[t]$ and $i,j\in[k]$.
The private envy-free condition gives
$$
\sum_g x_g^i v_i(g)\ge \sum_g x_g^j v_i(g)
$$
for all $i,j\in[k]$.

Finally, Theorem~\ref{thm:jojic} gives each $F_i$ as a union of $t+1$
intervals. Hence, for each color $i$, only $2(t+1)$ item intervals can be cut by
the boundary of $F_i$, so only $2(t+1)$ items have $0<x_g^i<1$. Thus $x$ is
$O(t)$-sparse.
\end{proof}

We can now prove the desired discrepancy upper bound.

\begin{theorem}\label{thm:discpe_upper}
Let \(k\ge 2\) and \(t\ge 1\).  Let
\(\mathcal{V}=\{v_1,\ldots,v_k\}\) and
\(\mathcal{U}=\{u_1,\ldots,u_t\}\) be additive valuations on \(m \geq k\) items, with
all item values in \([0,1]\).  If \(k\) is a prime power, then
\[
\discpe(\mathcal{V},\mathcal{U})
\le
O\!\left(
\min\left\{
t,\sqrt{t\log k}
\right\}
\right).
\]
For arbitrary \(k\),
\[
\discpe(\mathcal{V},\mathcal{U})
\le
O\!\left(
\min\left\{
t,\sqrt{t\log k}
\right\}
\cdot
\frac{\log m}{\log k}
\right).
\]
\end{theorem}
\begin{proof}
Let $
D=\min\{t,\sqrt{t\log k}\}.
$
First, by treating the private valuations as additional consensus valuations,
the standard multicolor discrepancy bound of~\citet{MulticolorDoerr} gives
$$
\discpe(\mathcal V,\mathcal U)
\le O(\sqrt{t+k}).
$$
Thus, if $t\ge k$, then
$\discpe(\mathcal V,\mathcal U)\le O(\sqrt{t+k}) \leq O(\sqrt t) \leq  O(D)$.
Hence we may assume $t\le k$. 

Assume first that $k$ is a prime power. By
Lemma~\ref{lem:sparse_fractional_allocation}, there is an $O(t)$-sparse
fractional allocation $x$ which is exactly balanced for every $u_\ell$ and
exactly envy-free for every $v_i$. Apply Lemma~\ref{lem:sparse_rounding} to
the valuations in $\mathcal U\cup\mathcal V$. Here $q=O(t)$ and $d=t+k$, so,
since $t\le k$, $
\log(kd)=O(\log k).
$

We obtain an integral allocation $\mathcal A=(A_1,\ldots,A_k)$ such
that, for every $w\in\mathcal U\cup\mathcal V$ and every $i\in[k]$,
$$
|w(A_i)-x_i(w)|\le O(D),
\qquad
x_i(w):=\sum_g x_g^i w(g).
$$

For any consensus valuation $u_\ell$, the fractional allocation satisfies
$x_i(u_\ell)=x_j(u_\ell)$. Hence
$$
\begin{aligned}
|u_\ell(A_i)-u_\ell(A_j)|
&\le |u_\ell(A_i)-x_i(u_\ell)|  \\
&\quad + |u_\ell(A_j)-x_j(u_\ell)| \\
&\le O(D).
\end{aligned}
$$
Similarly, for every private valuation $v_i$, we have
$x_i(v_i)\ge x_j(v_i)$. Thus
$$
\begin{aligned}
v_i(A_j)-v_i(A_i)
&\le |v_i(A_j)-x_j(v_i)| \\
&\quad + x_j(v_i)-x_i(v_i) \\
&\quad + |x_i(v_i)-v_i(A_i)| \\
&\le O(D).
\end{aligned}
$$
This proves the prime-power case.

Now let $k$ be a non-prime power, and let
$K\ge k$ be the smallest prime power at least $k$. Since every prime is a
prime power, the prime-gap bound of Baker--Harman--Pintz gives
$
K=k+O(k^{0.525}).
$ Thus
$$
\log K=\Theta(\log k),
\qquad
\alpha:=\frac{K-k}{K}=O(k^{-0.475}).
$$

We recursively allocate the items. At a step with remaining item set $M'$, apply
the prime-power case with $K$ colors to the private valuations
$
v_1,\ldots,v_k,0,\ldots,0
$
and to the consensus valuations $\mathcal U\cup\{\lambda\}$, where
$\lambda(g)=1$. This gives a $K$-allocation
$(C_1,\ldots,C_K)$ with error $O(D)$, since $t+1=O(t)$ and
$\log K=\Theta(\log k)$.

The bundles assigned to the first $k$ (real) colors
are permanently allocated, while the bundles assigned to the remaining
$K-k$ (dummy) colors constitute the unallocated remainder. Since $\lambda$ is
the cardinality valuation, the size of this remainder is at most a
$\alpha$-fraction of the current item set (up to the $O(D)$ discrepancy),
so the number of remaining items decreases geometrically with the recursion
depth. 

More formally add $C_i$ permanently to the final bundle of color $i$, for $i\in[k]$, and
recurse on
$$
M''=C_{k+1}\cup\cdots\cup C_K.
$$
Let $m'=|M'|$ and $m''=|M''|$. Since the cardinality valuation is balanced up
to $O(D)$, each dummy bundle has size at most
$
\frac{m'}{K}+O(D).
$
Therefore
$$
m''
\le
(K-k)\left(\frac{m'}{K}+O(D)\right)
=
\alpha m' + O((K-k)D).
$$
Unrolling the recurrence gives
\[
\begin{aligned}
m_r &\le \alpha^r m + O((K-k)D)\sum_{s=0}^{r-1}\alpha^s \le \alpha^r m + O(kD).
\end{aligned}
\]
Choose $r$ so that $\alpha^r m\le kD$. Since
$\log(1/\alpha)=\Omega(\log k)$, we stop after at most $r$ iterations with
$$
r=O\left(\frac{\log m}{\log k}\right).
$$

Allocate the remaining $O(kD)$ items arbitrarily, but as evenly as
possible, among the $k$ final bundles. Each bundle receives $O(D)$ additional
items, so this cleanup step contributes only $O(D)$ extra error.

Each recursive level contributes $O(D)$ to both the consensus discrepancy and
the private-envy error, and these errors add over the levels. Hence

$$
\discpe(\mathcal V,\mathcal U)
\le
O\left(
\min\{t,\sqrt{t\log k}\}\cdot
\frac{\log m}{\log k}
\right).
$$
\end{proof}

We finally derive Theorem~\ref{thm:mainub}.  Let \(f(k,t,m)\) be the upper bound
on \(\discpe(\mathcal{V},\mathcal{U})\) given by
Theorem~\ref{thm:discpe_upper}, rounded up to the nearest integer.  By the
second part of Lemma~\ref{lem:discrepancy_reduction}, the corresponding group
allocation instance admits an EF\(c\) allocation for
$
c= 4f(k,2t,m).
$
Since
\[
f(k,2t,m)
=
O\!\left(
\min\left\{
t,\sqrt{t\log k}
\right\}
\right)
\]
in the prime-power case, and the same expression multiplied by
\(\frac{\log m}{\log k}\) in the arbitrary-\(k\) case, this
proves Theorem~\ref{thm:mainub}.

\bibliographystyle{plainnat}
\bibliography{references}

\include{appendix}

\end{document}

%% file: appendix.tex
\appendix

\section{Missing proofs of the upper bound section}\label{app:proofs_UB}

\discrepancyreduction*

\begin{proof}
We first prove the lower-bound direction. Suppose, toward a contradiction,
that $\mathcal A=(A_1,\ldots,A_k)$ is EF$c$. Fix $i,j\in[k]$. Since the
private agent $v_i$ belongs to group $i$, there is a set $B\subseteq A_j$ with
$|B|\le c$ such that $v_i(A_i)\ge v_i(A_j\setminus B)$. As every item has
$v_i$-value at most $1$, we have
\[
v_i(A_j)-v_i(A_i)\le v_i(B)\le c.
\]
Thus $P(\mathcal V,\mathcal A)\le c$.

The same argument applies to the consensus valuations. For any
$\ell\in[t]$ and any $i,j\in[k]$, applying EF$c$ to the copy of $u_\ell$ in
group $i$ gives $u_\ell(A_j)-u_\ell(A_i)\le c$. Reversing the roles of $i$
and $j$ gives $u_\ell(A_i)-u_\ell(A_j)\le c$, and hence
$|u_\ell(A_i)-u_\ell(A_j)|\le c$. Therefore
$B(\mathcal U,\mathcal A)\le c$, so
\[
\discpe(\mathcal V,\mathcal U,\mathcal A)\le c.
\]
Minimizing over allocations yields $\discpe(\mathcal V,\mathcal U)\le c$,
contradicting the assumption that $\discpe(\mathcal V,\mathcal U)>c$.
Hence no EF$c$ allocation exists.

\medskip

We now prove the converse. Set $T:=f(k,2t,m)$. We begin with a short
``front-loading'' phase: run $T$ rounds of round robin on the private
valuations $v_1,\ldots,v_k$, in an arbitrary order. Whenever $v_i$ is called,
assign to group $i$ a remaining item of maximum $v_i$-value. Let $F_i$ be the
set of items assigned to group $i$ in this phase, and let
\[
R:=M\setminus \bigcup_{i=1}^k F_i
\]
be the set of remaining items.

The purpose of this phase is to give each private valuation a reserve of
high-value items that pay for the discrepancy error on the remaining
instance. This is the only difference from the proof of the upper bound
of~\citet{Manurangsi_ImprovedBounds}.

If the items are exhausted during the front-loading phase, then every bundle
has size at most $T$. Removing the entire target bundle therefore proves
EF$T$, and we are done. Hence assume the phase completes, so $|F_i|=T$ for
every $i$.

For each $i$, define $p_i:=\min_{g\in F_i} v_i(g)$. By the greedy choice
rule, every item in $R$ has $v_i$-value at most $p_i$, while
$v_i(F_i)\ge Tp_i$. Define a normalized private valuation on $R$ by
\[
\widehat v_i(g)=
\begin{cases}
v_i(g)/p_i, & p_i>0,\\
0, & p_i=0.
\end{cases}
\]
Then $\widehat v_i(g)\in[0,1]$ for every $g\in R$.

Next fix a consensus valuation $u_\ell$. Let
$L:=\min\{|R|,2Tk\}$, and let $S_\ell\subseteq R$ be a set of $L$ items of
largest $u_\ell$-value. Set
\[
q_\ell:=\min_{g\in S_\ell} u_\ell(g),
\]
with $q_\ell=0$ if $S_\ell=\emptyset$. We split the consensus valuation into
two normalized pieces: one that tracks the large items in $S_\ell$, and one
that tracks the remaining lower-valued tail. Namely, define
\[
y_\ell(g)=\mathbf 1[g\in S_\ell],
\qquad
z_\ell(g)=
\begin{cases}
u_\ell(g)/q_\ell, & g\in R\setminus S_\ell,\ q_\ell>0,\\
0, & \text{otherwise}.
\end{cases}
\]
Since every item outside $S_\ell$ has $u_\ell$-value at most $q_\ell$, both
$y_\ell$ and $z_\ell$ are $[0,1]$-valued.

Apply the definition of $T$ to the auxiliary instance on $R$ with private
valuations $\widehat v_1,\ldots,\widehat v_k$ and consensus valuations
\[
y_1,z_1,\ldots,y_t,z_t.
\]
If $|R|<m$, pad the instance with zero-value dummy items so that it has
exactly $m$ items, and then discard the dummy items from the resulting
partition. We obtain a partition $(R_1,\ldots,R_k)$ of $R$ with discrepancy
at most $T$. Finally, define
\[
A_i:=F_i\cup R_i
\qquad (i\in[k]).
\]

We first verify EF for the private agents. Fix $i,j\in[k]$ and remove $F_j$
from the target bundle. The auxiliary private-envy bound gives
\[
\widehat v_i(R_j)-\widehat v_i(R_i)\le T.
\]
Multiplying back, this yields $v_i(R_j)-v_i(R_i)\le Tp_i$; if $p_i=0$, the
same conclusion is immediate because $v_i$ vanishes on $R$. Since
$v_i(F_i)\ge Tp_i$, we get
\[
v_i(A_i)
=
v_i(F_i)+v_i(R_i)
\ge
v_i(R_j)
=
v_i(A_j\setminus F_j).
\]
Thus every private agent is EF$T$.

It remains to verify EF for the consensus agents. Fix $\ell\in[t]$ and
$i,j\in[k]$, and write
\[
S=S_\ell,\qquad q=q_\ell,\qquad y=y_\ell,\qquad z=z_\ell.
\]
We will remove from the target bundle the front-loaded items and the
high-value consensus items assigned to group $j$:
\[
B:=F_j\cup (R_j\cap S).
\]
Let $r_a:=|R_a\cap S|$ for $a\in[k]$. Since $y$ is balanced up to discrepancy
$T$, we have $|r_a-r_b|\le T$ for all $a,b\in[k]$. Also
$\sum_a r_a=|S|=L\le 2Tk$, so
\[
r_j\le \frac{L+(k-1)T}{k}\le 3T.
\]
Therefore
\[
|B|\le |F_j|+|R_j\cap S|\le T+3T=4T.
\]

If $S=R$, then $A_j\setminus B=\emptyset$, so there is nothing to prove.
Assume then that $S\ne R$. In this case $L=2Tk$. Using again
$|r_a-r_b|\le T$, we obtain
\[
r_i\ge \frac{L-(k-1)T}{k}\ge T.
\]
Every item in $S$ has $u_\ell$-value at least $q$, and hence
$u_\ell(R_i\cap S)\ge Tq$. Moreover, the balance of $z$ gives
\[
u_\ell(R_j\setminus S)-u_\ell(R_i\setminus S)\le Tq,
\]
with the case $q=0$ being immediate since all items outside $S$ then have
$u_\ell$-value $0$. Consequently,
\[
\begin{aligned}
u_\ell(A_i)
&\ge u_\ell(R_i) \\
&= u_\ell(R_i\cap S)+u_\ell(R_i\setminus S) \\
&\ge Tq+u_\ell(R_i\setminus S) \\
&\ge u_\ell(R_j\setminus S) \\
&= u_\ell(A_j\setminus B).
\end{aligned}
\]
Thus every consensus agent is envy-free after removing at most $4T$ items
from the target bundle.

We have constructed an EF$(4T)$ allocation. Since
$T=f(k,2t,m)$, this is an EF$c$ allocation for
\[
c=4f(k,2t,m).
\]
\end{proof}

\sparserounding*

\begin{proof}
For each item $g\in M$, independently choose a color $I(g)\in[k]$ with
$$
\Pr[I(g)=i]=x_g^i,
$$
and put $g$ in the bundle $A_{I(g)}$. Since colors with $x_g^i=0$ are never
chosen, the rounding is support-preserving.

Fix $\ell\in[d]$ and $i\in[k]$. Let
$$
S_i=\{g\in M:1 > x_g^i>0\},
\qquad
\mu_{\ell i}=\sum_{g\in M}x_g^i w_\ell(g).
$$
By $q$-sparsity, $|S_i|\le q$. Moreover, items with either $x_g^i = 0$ or $x_g^i = 1$ are rounded deterministically and do not contribute to the error of the rounding. Therefore,
$$
w_\ell(A_i)-\mu_{\ell i}
=
\sum_{g\in S_i} X_g^{\ell i},
$$
where
$$
X_g^{\ell i}
=
\bigl(\mathbf 1_{\{I(g)=i\}}-x_g^i\bigr)w_\ell(g).
$$
The variables $X_g^{\ell i}$ are independent, have mean zero, and lie in
$[-1,1]$. Hence Hoeffding's inequality gives, for every $\tau>0$,
$$
\Pr\bigl[|w_\ell(A_i)-\mu_{\ell i}|>\tau\bigr]
\le
2\exp\left(-\frac{\tau^2}{2q}\right).
$$

Let
$$
E_{\ell i}
=
\bigl\{|w_\ell(A_i)-\mu_{\ell i}|>\tau\bigr\}.
$$
By the union bound,
$$
\begin{aligned}
\Pr\left[\bigcup_{\ell=1}^d\bigcup_{i=1}^k E_{\ell i}\right]
&\le
\sum_{\ell=1}^d\sum_{i=1}^k \Pr[E_{\ell i}]  \\
&\le
2kd\exp\left(-\frac{\tau^2}{2q}\right).
\end{aligned}
$$
Now take $\tau=C\sqrt{q\log(kd)}$. Then the last bound is
$$
2kd\exp\left(-\frac{C^2\log(kd)}{2}\right)
=
2(kd)^{1-C^2/2}.
$$
Choosing $C$ as a sufficiently large universal constant makes this quantity
less than $1$. Therefore, with positive probability,
$$
|w_\ell(A_i)-\mu_{\ell i}|
\le
C\sqrt{q\log(kd)}
$$
holds simultaneously for all $\ell\in[d]$ and $i\in[k]$.

Also, every support-preserving rounding satisfies
$$
\begin{aligned}
|w_\ell(A_i)-\mu_{\ell i}|
&\le
\sum_{g\in S_i}
\bigl|\mathbf 1_{\{I(g)=i\}}-x_g^i\bigr|w_\ell(g) \\
&\le
\sum_{g\in S_i} w_\ell(g)
\le q,
\end{aligned}
$$
since $w_\ell(g)\in[0,1]$ and $|S_i|\le q$. Thus the rounding above satisfies
both estimates. 
\end{proof}